\documentclass[12pt]{article}
\usepackage{latexsym}
\usepackage{graphicx}
\title{Hidden variables with nonlocal time}
\author{Hrvoje Nikoli\'c \\
Theoretical Physics Division, Rudjer Bo\v{s}kovi\'{c} Institute, \\
P.O.B. 180, HR-10002 Zagreb, Croatia \\
{\normalsize e-mail: hrvoje@thphys.irb.hr} \\
\makebox[1in]{} \\
}
\date{\today}
\begin{document}
\maketitle
\begin{abstract}
To relax the apparent tension between 
nonlocal hidden variables and relativity, we propose
that the observable proper time is not the same quantity as 
the usual proper-time parameter appearing in local relativistic equations. Instead,
the two proper times are related by a nonlocal rescaling parameter proportional
to $|\psi|^2$, so that they coincide in the classical limit. 
In this way particle trajectories may obey local
relativistic equations of motion in a manner consistent with the 
appearance of nonlocal quantum correlations. 
To illustrate the main idea, we first present two simple toy models
of local particle trajectories with nonlocal time,
which reproduce some nonlocal quantum phenomena.
After that, we present a realistic theory  
with a capacity to reproduce all predictions of quantum theory.
\end{abstract}


\maketitle

\section{Introduction}

According to the Bell \cite{bell} and other \cite{ghz,hardy} theorems, 
any hypothetic hidden-variable completion of quantum mechanics (QM) must 
necessarily be nonlocal. Yet, the theorem has little to say about the
details of such a hypothetic hidden-variable theory. To actually find 
a promissing theory of that kind,
one has to propose something that cannot be derived from already known facts.

The best known theory of that kind is nonrelativistic Bohmian mechanics \cite{bell,bohm,BMbook1}.
In this theory, the velocity of each particle at a given time depends on the 
positions of all other particles at the same time. However, such a theory is not 
relativistic-covariant because the notion of ``same time'' depends on the choice
of spacetime coordinates. To overcome this problem,
a relativistic-covariant version of Bohmian mechanics
has been developed, in which the usual space probability density is generalized to the
spacetime probability density \cite{nikbosfer,nikijqi,hern,nikqft,niktorino,nikSmatr}.
In this relativistic theory, the notion of ``same time'' is replaced by a notion of 
``same scalar parameter'', which is a parameter similar to proper time 
in classical relativistic mechanics. In \cite{niksuperlum}
this parameter is interpreted as a generalized proper time.  
In the present paper we refer to this parameter as {\em modified proper time},
with a motivation to further refine the physical meaning of it.
 
Aesthetically, there is one unappealing feature of relativistic Bohmian mechanics.
While nonrelativistic Bohmian mechanics involves essentially only one
unconventional idea -- a nonlocal law for particle velocities, the relativistic variant
involves two unconventional ideas -- a nonlocal law for particle velocities {\em and}
the concept of modified proper time.
To reduce the number of unconventional ideas involved in a relativistic-covariant
nonlocal hidden-variable theory, in this paper we propose a new theory in which the modified
proper time is {\em the only} unconventional idea. In our proposal, 
particle trajectories satisfy
{\em local} relativistic equations of motion, but the 
usual relativistic proper time associated with each of the trajectories is {\em not} the
true physical time. Instead, the true physical time is the modified proper time
nonlocally related to the usual proper time. The dynamics reparameterized
in terms of this new time looks nonlocal, in a manner compatible with predictions of QM. 

The physical idea lying behind this construction can be understood in simple 
nonrelativistic terms as follows. Consider $n$ particles (in 3 space dimensions)
with velocities $v^i_a$, $i=1,2,3$, $a=1,\ldots,n$, and think of velocities
as ratios $v^i_a=\Delta x^i_a / \Delta t$ with small but finite $\Delta x^i_a$, $\Delta t$.
Clearly, the velocities obeying classical equations of motion 
 are not compatible with predictions of QM. Therefore, the velocities should be modified.
But mathematically, there are {\em two} ways to modify a velocity. The first way is to modify 
$\Delta x^i_a$ for the same $\Delta t$. Indeed, this is essentially what 
nonrelativistic Bohmian mechanics does. The second way is to modify
$\Delta t$ for the same $\Delta x^i_a$. This second way is much more restrictive
because it leaves the ratios $v^i_a/v^j_b$ unchanged. Yet, it may be nontrivial
if the ratio $\Delta t'/\Delta t$ between the new time $\Delta t'$ and the old time 
$\Delta t$ is not a constant, but a function of ${\bf x}_1,\ldots, {\bf x}_n$
(where ${\bf x}\equiv \{ x^1,x^2,x^3 \}$). We shall see that by an appropriate choice
of such a nonlocal function, it is possible to get new ``apparent'' velocities
$\Delta x^i_a / \Delta t'$ that reproduce some predictions of QM.
Moreover, we shall see that a relativistic variant of that idea may reproduce 
all QM predictions.

To provide a gentle exposition of our idea, we split it into three 
conceptually independent steps. In the first step (Sec.~\ref{SECnrtoy})
we present a nonrelativistic toy model based on classical
trajectories of free particles. We explain how appropriate nonlocal rescaling 
of the nonrelativistic time may mimic some nonlocal quantum phenomena
in nonrelativistic QM. In the second step (Sec.~\ref{SECrtoy})
we present a relativistic generalization of the idea in Sec.~\ref{SECnrtoy}.
This relativistic toy model, the main novelty of which (with respect to Sec.~\ref{SECnrtoy})
is the introduction of the concept of modified proper time,
remedies many (but not all) deficiencies of the nonrelativistic model.
In the third step (Sec.~\ref{SECreal}) we present a fully realistic
theory in which particle trajectories in spacetime obey local, 
but non-classical equations of motion. In Sec.~\ref{SECdisc} we discuss
our results and sketch how a theory of that form 
may be further generalized to reproduce all predictions of quantum theory.

\section{Nonrelativistic toy model}
\label{SECnrtoy}

Consider $n$ free nonrelativistic particles obeying classical laws of motion
described by the Hamilton-Jacobi equation
\begin{equation}\label{HJclas}
 \sum_{a=1}^{n} \frac{(\nabla_a {\cal S})^2 }{2m_a} = -\frac{\partial {\cal S} }{\partial t} .
\end{equation}
The solution of  (\ref{HJclas}) is
\begin{equation}
{\cal S}({\bf x}_1,\ldots, {\bf x}_n,t)=S({\bf x}_1,\ldots, {\bf x}_n)-Et ,
\end{equation}
where
\begin{equation}
S({\bf x}_1,\ldots, {\bf x}_n)=\sum_{a=1}^{n} {\bf p}_a \cdot {\bf x}_a ,
\end{equation}
\begin{equation}
 E=\sum_{a=1}^{n} \frac{{\bf p}^2_a }{2m_a} ,
\end{equation}
and ${\bf p}_a$ are the particle momenta, which are constants of motion. 
The velocities of particles are
\begin{equation}\label{e5}
 \frac{d{\bf X}_a(t)}{dt} \stackrel{\rm traj}{=} {\bf v}_a({\bf X}_a(t)) ,
\end{equation}
where $\stackrel{\rm traj}{=}$ denotes equalities valid only along physical particle trajectories
and
\begin{equation}\label{e6}
 {\bf v}_a({\bf x}_a) \equiv \frac{\nabla_a S({\bf x}_1,\ldots, {\bf x}_n)}{m_a} 
= \frac{{\bf p}_a}{m_a}.
\end{equation}
Here ${\bf v}_a({\bf x}_a)$ is a constant function, but the notation ${\bf v}_a({\bf x}_a)$
reminds us that it would be a non-constant function of ${\bf x}_a$
if local forces on particles (caused by a background potential) were present. 

Eq.~(\ref{e6}) shows that, for free particles, the velocity function is divergence-free, i.e., 
\begin{equation}\label{e7}
 \sum_{a=1}^{n} \nabla_a {\bf v}_a = 0.
\end{equation}
For the reasons which will become clear later, we write this as
\begin{equation}\label{e8}
 \sum_{a=1}^{n} \nabla_a (\rho {\bf v}'_a ) =0 ,
\end{equation}
where $\rho({\bf x}_1,\ldots, {\bf x}_n)$ is a function which we shall specify later and
\begin{equation}\label{e9}
 {\bf v}'_a({\bf x}_1,\ldots, {\bf x}_n) \equiv 
\frac{{\bf v}_a({\bf x}_a)}{\rho({\bf x}_1,\ldots, {\bf x}_n)} . 
\end{equation}

What is the  physical interpretation of the quantity ${\bf v}'_a$ defined by (\ref{e9})?
Before answering that question, let us first explore some mathematical features.
Mathematically, along the trajectories (\ref{e5}), ${\bf v}'_a$ 
can be thought of as the particle velocity
defined with respect to a rescaled time $t'\neq t$. Indeed, to find the relation between
$t$ and $t'$, we write (\ref{e9}) along the trajectories as
\begin{equation}
 \frac{d{\bf x}_a}{dt'} \stackrel{\rm traj}{=} \frac{1}{\rho} \frac{d{\bf x}_a}{dt} .
\end{equation}
From this we see that, along the trajectories, $t$ and $t'$
must be related as $dt' \stackrel{\rm traj}{=} \rho \, dt$.
We, however, are allowed to propose a more general relation
\begin{equation}\label{e11}
dt'=\rho({\bf x}_1,\ldots, {\bf x}_n) \, dt ,
\end{equation} 
which is a coordinate transformation in the $(3n+1)$-dimensional space
(with a trivial transformation of space coordinates ${\bf x}'_a={\bf x}_a$) valid everywhere,
not only along the trajectories.
(For further clarifications, see also the Appendix.)
This means that the integrated form of (\ref{e11}) is
$t'=\rho({\bf x}_1,\ldots, {\bf x}_n) \, t + {\rm const}$, but the value of
${\rm const}$ will be irrelevant in further discussions. (To provide that 
$t$ and $t'$ increase in the same direction
along any trajectory, $\rho({\bf x}_1,\ldots, {\bf x}_n)$
must be a non-negative function.)
Thus, the classical trajectories (\ref{e5}) can also be thought of as solutions
of the {\em nonlocal} equations
\begin{equation}\label{e12}
 \frac{d\tilde{{\bf X}}_a(t')}{dt'} \stackrel{\rm traj}{=} 
{\bf v}'_a(\tilde{{\bf X}}_1(t'),\ldots, \tilde{{\bf X}}_n(t')) ,
\end{equation}
where $\tilde{{\bf X}}_a(t') 
\stackrel{\rm traj}{\equiv} {\bf X}_a(t(t'))$ for $t(t')$ evaluated along the
trajectories. 
In this sense, {\em the local law (\ref{e5}) is mathematically equivalent
to the nonlocal law (\ref{e12})}. 
Furthermore, since
$\rho({\bf x}_1,\ldots, {\bf x}_n)$ does not have an explicit dependence on
$t'$, so that $\partial \rho/\partial t' =0$,
we see that (\ref{e8}) can also be written as a continuity equation
\begin{equation}\label{e13}
\frac{\partial\rho}{\partial t'} + \sum_{a=1}^{n} \nabla_a (\rho {\bf v}'_a ) =0 .
\end{equation}

Now we are ready for the physical interpretation. We fix
\begin{equation}\label{e14}
 \rho({\bf x}_1,\ldots, {\bf x}_n) = |\psi({\bf x}_1,\ldots, {\bf x}_n)|^2 ,
\end{equation}
where $\psi$ is the {\em quantum-mechanical} $n$-particle wave function.
Then the particle trajectories (\ref{e12}) are compatible 
with the quantum probabilistic distribution of particles (\ref{e14}). 
Namely, if a statistical ensemble of particles has the distribution (\ref{e14})
for some initial time $t'_0$, then (\ref{e13}) and (\ref{e12}) imply that
the ensemble will have the distribution (\ref{e14}) for {\em any} time
$t'$. Indeed, this is very similar to the Bohmian interpretation of
QM, where a nonlocal law similar to (\ref{e12}), together with a continuity 
equation of the form of (\ref{e13}), also provides the consistency 
with QM.
 
Nevertheless, there are several important differences with respect to 
nonrelativistic Bohmian mechanics. First, in the present theory
the equations for particle velocities can be written in a {\em local} form
(\ref{e5}). Nothing like that is possible in Bohmian mechanics for a
nonlocal distribution $|\psi({\bf x}_1,\ldots, {\bf x}_n)|^2$ (where, by a nonlocal distribution, 
we mean a distribution that cannot be written as a local product 
$|\psi_1({\bf x}_1) \cdots \psi_n({\bf x}_n)|^2$).

Second, in Bohmian mechanics, the quantum probability is conserved
with respect to time $t$. By contrast, owing to (\ref{e13}), 
in the present theory the quantum probability is conserved
with respect to a new time $t'$. Thus, if we want 
(\ref{e13}) to represent a {\em physical} probability conservation,
then $t'$ must represent a {\em physical} time. This suggests
the following provisional physical idea: {\em The physically measurable time is not
$t$, but $t'$}. In other words, clocks at different positions in space
run at different rates, which implies that constant
velocities given by (\ref{e5})-(\ref{e6}) appear as non-constant velocities
given by (\ref{e12}). Conceptually, this is similar to the position-dependent 
running of clocks in the general theory of relativity, but with an important difference
that general relativity is a local theory, while (\ref{e11}) is a 
{\em nonlocal} law. (See also the Appendix for further clarifications.) 
If we imagine that each particle has its own clock
that measures time $t'$, then the running of each clock depends not only on the position of that clock, but also on the positions of {\em all other clocks}.

As a consistency check, note also that $t'=t$ in the classical limit.
The simplest way to see this is to consider a pure-phase wave function
of the form $\psi({\bf x}_1, \ldots, {\bf x}_n,t)=
e^{i {\cal S}({\bf x}_1, \ldots, {\bf x}_n,t)/\hbar}$, for some real function
${\cal S}$. It is well-known that the Schr\"odinger equation for such a
wave function reduces to the classical Hamilton-Jacobi equation.
Therefore, the wave function in the classical limit is a 
pure-phase wave function, which means that $|\psi|^2$ is a constant.
Consequently, (\ref{e11})  in the classical limit becomes
\begin{equation}
 dt'={\rm const} \, dt ,
\end{equation}
so, up to the irrelevant choice of units, $t'$ coincides with $t$.
 
This suggests that a nonlocal time could be {\em the origin} of 
quantum nonlocalities. Yet, at the moment it would be premature 
to claim that such an idea is viable. Namely, the third difference
with respect to Bohmian mechanics is the fact that the idea
above works only when $|\psi({\bf x}_1,\ldots, {\bf x}_n)|^2$
does not have an explicit dependence on time. This is valid
only for energy-eigenstates $\psi$'s, which is far from being 
the most general case. Besides, the theory presented above
is not relativistic. In addition, it works only for free particles (which, however,
may be entangled) because (\ref{e7}) is no longer valid in the interacting case. 
Thus, the model presented in this section is only a toy model.
To remedy its deficiencies the idea should be
further developed, which we do in the subsequent sections.

\section{Relativistic toy model}
\label{SECrtoy}

In this section we present a relativistic version of the ideas introduced
in Sec.~\ref{SECnrtoy}. 
Let $x=\{x^{\mu}\}$, $\mu=0,1,2,3$, denotes the coordinates
of a position in spacetime, with the metric signature $(+,-,-,-)$.
The relativistic Hamilton-Jacobi equation for $n$
free relativistic particles is (see, e.g., \cite{niksuperlum}) 
\begin{equation}\label{HJclasr}
 -\sum_{a=1}^{n} \frac{(\partial_a^{\mu} S) (\partial_{a\mu} S) }{2m_a} 
+\sum_{a=1}^{n} \frac{m_a}{2}= 0,
\end{equation}
where $\partial_{a\mu}\equiv \partial/\partial x^{\mu}_a$.
(The Einstein convention of summation over repeated indices refers
only to vector indices $\mu$, while the summation over the particle labels $a$
is to be performed only when the summation $\sum_a$ is indicated explicitly.)
The solution of (\ref{HJclasr}) is
\begin{equation}
S(x_1,\ldots, x_n)=-\sum_{a=1}^{n} p_{a\mu}x_a^{\mu} ,
\end{equation}
where the constants of motion $p^{\mu}_a$ are the particle 4-momenta satisfying
the mass-shell condition 
\begin{equation}\label{massshell}
 p^{\mu}_a p_{a\mu} = m_a^2 .
\end{equation}
The 4-velocities of particles are
\begin{equation}\label{e5r}
 \frac{dX^{\mu}_a(s)}{ds} \stackrel{\rm traj}{=} v^{\mu}_a(X_a(s)) ,
\end{equation}
where
\begin{equation}\label{e6r}
v^{\mu} _a(x_a) \equiv -\frac{\partial^{\mu}_a S(x_1,\ldots, x_n)}{m_a} 
= \frac{p^{\mu}_a}{m_a} ,
\end{equation}
and $s$ is a scalar parameter that parameterizes the trajectories. 
Eq.~(\ref{massshell}) implies that (\ref{e6r}) are unit velocities $v^{\mu}_a v_{a\mu}=1$, 
so (\ref{e5r}) implies
\begin{equation}
 dX^{\mu}_a dX_{a\mu} \stackrel{\rm traj}{=} ds^2.
\end{equation}
This shows that $s$ is not just any parameter along the trajectory, but is equal to the proper time
along the trajectory. In classical relativistic mechanics, the proper time (and not the
coordinate time $x^0$) is the physical time.

Eq.~(\ref{e6r}) implies 
\begin{equation}\label{e7r}
 \sum_{a=1}^{n} \partial_{a\mu} v^{\mu}_a = 0 .
\end{equation}
This can also be written as
\begin{equation}\label{e8r}
 \sum_{a=1}^{n} \partial_{a\mu} (\rho v'^{\mu}_a) =0 ,
\end{equation}
where $\rho(x_1,\ldots, x_n)$ is a function to be specified later and
\begin{equation}\label{e9r}
 v'^{\mu}_a(x_1,\ldots, x_n) \equiv 
\frac{v^{\mu}_a(x_a)}{\rho(x_1,\ldots, x_n)} . 
\end{equation}
Therefore, in analogy with the results of Sec.~\ref{SECnrtoy}, $v'^{\mu}_a$ can be interpreted
as the velocity calculated with respect to a new scalar parameter $s'$, 
where the relation between $s'$ and $s$ is given by
\begin{equation}\label{e11r}
ds'=\rho(x_1,\ldots, x_n) \, ds .
\end{equation}
This is a coordinate transformation in the $(4n+1)$-dimensional space, which means
that (analogously to the parameters $t$ and $t'$ in Sec.~\ref{SECnrtoy}) 
$s$ and $s'$ are viewed as parameters existing even without the trajectories.
(See the Appendix for further clarifications.)
In this way, the classical local relativistic law of motion (\ref{e5r}) is mathematically equivalent
to the {\em nonlocal relativistic} law 
\begin{equation}\label{e12r}
 \frac{d\tilde{X}^{\mu}_a(s')}{ds'} \stackrel{\rm traj}{=} v'^{\mu}_a (\tilde{X}_1(s'),\ldots, \tilde{X}_n(s')) ,
\end{equation}
where $\tilde{X}^{\mu}_a(s') 
\stackrel{\rm traj}{\equiv} X^{\mu}_a(s(s'))$ for $s(s')$ evaluated along the
trajectories. 
Indeed, from (\ref{e5r}), (\ref{e9r}) and (\ref{e12r}) we see that,
along the trajectories, the following parameterization-independent equalities are valid
\begin{equation}\label{param2'}
 \frac{dx^{\mu}_a}{dx^{\nu}_b} \stackrel{\rm traj}{=} \frac{v^{\mu}_a(x_a)}{v^{\nu}_b(x_b)} \stackrel{\rm traj}{=}
\frac{v'^{\mu}_a(x_1,\ldots,x_n)}{v'^{\nu}_b(x_1,\ldots,x_n)} ,
\end{equation}
which reflects the fact that (\ref{e5r}) and (\ref{e12r}) correspond to two different parameterizations of {\em the same trajectories in spacetime}.
Eq.~(\ref{e12r}) can be viewed as a nonlocal parameterization of the local trajectories
(\ref{e5r}). 
Furthermore, since
$\rho(x_1,\ldots, x_n)$ does not have an explicit dependence on
$s'$, we see that (\ref{e8r}) can also be written as 
\begin{equation}\label{e13r}
\frac{\partial\rho}{\partial s'} +  \sum_{a=1}^{n} \partial_{a\mu} (\rho v'^{\mu}_a)=0 .
\end{equation}

These formal manipulations become physically interesting when we fix
\begin{equation}\label{e14r}
 \rho(x_1,\ldots, x_n) = \frac{ |\psi(x_1,\ldots, x_n)|^2 }{N} ,
\end{equation}
where $N$ is a normalization constant to be fixed later and
$\psi(x_1,\ldots, x_n)$ is the relativistic many-time wave function satisfying
$n$ Klein-Gordon equations
\begin{equation}\label{KG}
(\partial_a^{\mu}\partial_{a\mu} +m_a^2)\psi(x_1,\ldots,x_n)=0,
\end{equation}
one for each $x_a$. (Here and in the rest of the paper we use units $\hbar=1$.)
Eq.~(\ref{e13r}) has the form of a relativistic equivariance equation, analogous to that
in relativistic Bohmian mechanics \cite{BMbook1,nikbosfer,nikijqi,hern,nikqft,niktorino,nikSmatr}.
Thus, the trajectories (\ref{e12r}) are compatible with statistical predictions of the ``standard'' purely probabilistic interpretation of QM.  
Namely, if a statistical ensemble of particles has the probability distribution (\ref{e14r})
(on the relativistic $4n$-dimensional configuration space)
for some initial $s'$, then the equivariance equation (\ref{e13r}) provides that
the ensemble will have the distribution (\ref{e14r}) for {\em any} $s'$.

Of course, similarly to the nonrelativistic case in Sec.~\ref{SECnrtoy}, 
here the essential physical assumption is the idea that physically observable time is $s'$
given by (\ref{e11r}), and not the usual classical relativistic proper time $s$.
In this paper we do not attempt to explain why $s'$ should be more physical than $s$.
We simply {\em assume} that it is. Such an assumption may seem {\em ad hoc},
but we find it remarkable that such a simple assumption alone may reproduce
the predictions of QM, practically without any other unconventional assumptions.
In addition, analogously to the nonrelativistic case in Sec.~\ref{SECnrtoy},
one can show that (\ref{e11r}) with (\ref{e14r}) implies that $s'=s$ in the classical limit, 
which demonstrates the consistency of our assumption.
Thus, the modified proper time $s'$ can also be thought of as the quantum proper 
time.

In fact, the whole idea of this section is very similar to that of Sec.~\ref{SECnrtoy}.
Yet, unlike Sec.~\ref{SECnrtoy}, the present section describes a model which is 
relativistic covariant and is not restricted to energy eigenstates. 
Therefore, this relativistic model seems much more promissing than the
nonrelativistic one. Nevertheless, the present model is still only a toy model
because it describes only free particles, not particles with interaction.
A theory that can deal with interactions will be presented in the next section.

Let us end this section with two additional remarks. First,
the spacetime
probabilistic interpretation of (\ref{e14r}) in the $4n$-dimensional configuration
space implies that $N$ should be fixed to
\begin{equation}
 N=\int d^4x_1\cdots d^4x_n |\psi(x_1,\ldots,x_n)|^2 .
\end{equation}
To avoid dealing with an infinite $N$, one can confine the whole physical system into
a large but finite 4-dimensional spacetime box. 
There are also mathematically more rigorous ways 
of dealing with wave functions that do not vanish at infinity,
such as the rigged Hilbert space \cite{ballentine}.

Second, the spacetime probabilistic interpretation of (\ref{e14r}), generalizing the usual 
space probabilistic interpretation of nonrelativistic QM, has also been
studied in older literature, such as \cite{stuc,broyles}. A detailed discussion of 
compatibility of such a generalized probabilistic interpretation
with the usual probabilistic interpretation is presented in \cite{niktorino}.
In particular, in \cite{niktorino} it is explained how particle trajectories 
obeying (\ref{e13r}) are compatible with {\em all} statistical predictions
of QM, not only with statistical predictions on particle positions.
The key insight is that all observations can be reduced
to observations of spacetime positions of some macroscopic pointer observables. 

\section{A realistic theory}
\label{SECreal}

Eq.~(\ref{e7r}) is not valid for classical trajectories with interactions.
To incorporate interactions, in
this section we propose a different local law for particle trajectories.
These will be nonclassical trajectories determined by the wave function
in a manner similar to that in Bohmian mechanics, but with an important
difference that trajectories in Bohmian mechanics obey a nonlocal law.

\subsection{Free particles}
\label{SECfree}

From the wave function satisfying (\ref{KG}),
one can construct the quantity
\begin{equation}\label{curn}
j_{\mu_1\ldots\mu_n}(x_1,\ldots,x_n) \equiv \left( \frac{i}{2} \right)^n \psi^*
\!\stackrel{\leftrightarrow}{\partial}_{\mu_1}\! \cdots
\!\stackrel{\leftrightarrow}{\partial}_{\mu_n}\! \psi ,
\end{equation}
where 
\begin{equation}\label{antisimder}
\chi \!\stackrel{\leftrightarrow\;}{\partial_{\mu}}\! \varphi \equiv
\chi\partial_{\mu}\varphi - (\partial_{\mu}\chi)\varphi ,
\end{equation} 
and $\partial_{\mu_a}\equiv \partial/\partial x^{\mu_a}_a$.
The quantity (\ref{curn})
transforms as an $n$-vector \cite{witt}. Eq.~(\ref{KG})  implies that 
this quantity satisfies
the conservation equation $\partial_{\mu_1}j^{\mu_1\ldots\mu_n}=0$
and similar conservation equations with other $\partial_{\mu_a}$.
Thus we have $n$ conservation equations
\begin{equation}\label{consn}
 \partial_{\mu_a}j^{\mu_1\ldots \mu_a \ldots \mu_n}=0 ,
\end{equation}
one for each $x_a$.
Assuming that $\psi$ is a superposition of positive-frequency solutions to (\ref{KG}),
$\psi$ can be normalized such that the $n$-particle Klein-Gordon norm is equal to 1.
Explicitly, this means that
\begin{equation}\label{globconsn}
\int_{\Sigma_1} dS^{\mu_1}_1 \cdots \int_{\Sigma_n} dS^{\mu_n}_n \,
j_{\mu_1\ldots\mu_n} =1,
\end{equation}
where $\Sigma_a$ are arbitrary 3-dimensional spacelike hypersurfaces and
\begin{equation}
dS_a^{\mu_a}=d^3x_a |g_a^{(3)}|^{1/2} n^{\mu_a}
\end{equation}
is the covariant measure of the 3-volume on $\Sigma_a$.
Here $n^{\mu_a}$ is the unit future-oriented vector normal to $\Sigma_a$,
while $g_a^{(3)}$ is the determinant of the induced metric on $\Sigma_a$.
The conservation equations (\ref{consn}) imply that
the left-hand side of (\ref{globconsn})
does not depend on the choice of spacelike hypersurfaces
$\Sigma_1,\dots,\Sigma_n$.

Now we introduce $n$ 1-particle currents
$j_{a\mu}(x_a)$ by omitting the integration over $dS^{\mu_a}_a$ in
(\ref{globconsn}). For example, for $a=1$,
\begin{equation}\label{eq6}
j_{1\mu}(x_1)=
\int_{\Sigma_2} dS^{\mu_2}_2 \cdots \int_{\Sigma_n} dS^{\mu_n}_n \,
j_{\mu \mu_2\ldots\mu_n}(x_1,\ldots,x_n) ,
\end{equation}
which does not depend on the choice of spacelike hypersurfaces
$\Sigma_2,\dots,\Sigma_n$
and satisfies $\partial_{1\mu} j_1^{\mu}=0$.
This implies the conservation equation
\begin{equation}\label{cons2}
 \sum_{a=1}^{n} \partial_{a\mu}v_a^{\mu}(x_a)=0 ,
\end{equation}
where
\begin{equation}\label{v=j/m}
v_a^{\mu}(x_a)=\frac{j_a^{\mu}(x_a)}{m_a} .
\end{equation}

Once we have the conservation equation (\ref{cons2}), we can proceed in the same way as in
Sec.~\ref{SECrtoy}. We propose that $v_a^{\mu}(x_a)$ are the particle velocities, i.e., 
that
\begin{equation}\label{param1}
 \frac{dX^{\mu}_a(s)}{ds} \stackrel{\rm traj}{=} v_a^{\mu}(X_a(s)) .
\end{equation}
(In \cite{nikfol}, the curves $X^{\mu}_a(s)$ have also been used as an auxiliary
mathematical tool.)
The spacetime trajectories obeying the local law (\ref{param1}) are the same as
the spacetime trajectories obeying the nonlocal law
\begin{equation}\label{param3}
 \frac{d\tilde{X}^{\mu}_a(s')}{ds'} \stackrel{\rm traj}{=} v'^{\mu}_a(\tilde{X}_1(s'),\ldots,\tilde{X}_n(s')) ,
\end{equation}
where 
\begin{equation}\label{e11ragain}
ds'=\rho(x_1,\ldots, x_n) \, ds 
\end{equation}
as in (\ref{e11r}), $\rho$ is given by (\ref{e14r}), and
\begin{equation}\label{eq10}
v'^{\mu}_a(x_1,\ldots,x_n) \equiv \frac{v^{\mu}_a(x_a)}{|\psi(x_1,\ldots,x_n)|^2} .
\end{equation}
The conservation equation (\ref{cons2}) now can be written as (\ref{e8r}), 
which leads to the equivariance equation (\ref{e13r}). In other words,
(\ref{cons2}) implies that the trajectories (\ref{param1}) are compatible with 
statistical predictions of QM, provided that the physical time is $s'$ given by
(\ref{e11ragain}).
In particular, similarly to the case of relativistic Bohmian mechanics \cite{niktorino},
even though (\ref{param1}) may lead to superluminal velocities,
a measured velocity cannot be superluminal.

Note that, in general, the parameter $s$ is not exactly equal to the proper time,
but is locally related to it. Indeed, from (\ref{param1}) and (\ref{v=j/m}) one finds
\begin{equation}\label{noproptime}
 ds^2 \stackrel{\rm traj}{=} f_a(X_a(s)) \,  dX^{\mu}_a dX_{a\mu} ,
\end{equation}
where $dX^{\mu}_a dX_{a\mu}$ is the squared proper-time interval and
\begin{equation}\label{f}
f_a(x_a) \equiv \frac{m_a^2}{j_a^{\mu}(x_a)j_{a\mu}(x_a)} 
\end{equation}
is a local function depending only on $x_a$.

Note also that, in the classical limit, the Klein-Gordon equations (\ref{KG}) reduce 
to the classical relativistic Hamilton-Jacobi equation (\ref{HJclasr}).
Consequently, by evaluating (\ref{eq6}) in that limit, it can be shown that  
(\ref{param1}) reduces to the classical relativistic equation of motion.
In particular, $s$ becomes exactly equal to the proper time. We shall 
show this more explicitly, and in a more general context, in Sec.~\ref{SECint}.

\subsection{Inclusion of interactions}
\label{SECint}

It is straightforward to generalize the theory in Sec.~\ref{SECfree} to particles
interacting with a classical gravitational or electromagnetic background,
by replacing the derivatives $\partial_{\mu}$ with the appropriate
covariant derivatives. Again, it can be shown that
the resulting theory has the correct classical limit. 

Let us see in more detail how it works for electromagnetic interactions.
The effects of the electromagnetic background $A^{\mu}(x)$ on the 
wave function of the particle with charge
$e$ are described by the gauge-covariant derivative
\begin{equation}
 D_{\mu}=\partial_{\mu}+ieA_{\mu}(x) .
\end{equation}
The appropriate generalization of (\ref{antisimder}) is \cite{nikcovpart}
\begin{equation}\label{antisimder2}
\chi \!\stackrel{\leftrightarrow\;}{D_{\mu}}\! \varphi \equiv
\chi D_{\mu}\varphi - (D^*_{\mu}\chi)\varphi ,
\end{equation}
where $D^*_{\mu}=\partial_{\mu}-ieA_{\mu}$.
Thus, the Klein-Gordon equations (\ref{KG}) generalize to
\begin{equation}\label{KGA}
(D_a^{\mu} D_{a\mu} +m_a^2)\psi(x_1,\ldots,x_n)=0,
\end{equation}
where
\begin{equation}
 D_{a\mu}=\partial_{a\mu}+ie_aA_{\mu}(x_a) .
\end{equation}
The quantity (\ref{curn}) generalizes to
\begin{equation}\label{curn2}
j_{\mu_1\ldots\mu_n}(x_1,\ldots,x_n) \equiv \left( \frac{i}{2} \right)^n \psi^*
\!\stackrel{\leftrightarrow}{D}_{\mu_1}\! \cdots
\!\stackrel{\leftrightarrow}{D}_{\mu_n}\! \psi ,
\end{equation}
where $D_{\mu_a}\equiv D_{a\mu_a}$.
Eq.~(\ref{KGA}) implies that the quantity (\ref{curn2}) satisfies
(\ref{consn})
 \begin{equation}\label{consnagain}
 \partial_{\mu_a}j^{\mu_1\ldots \mu_a \ldots \mu_n}=0 ,
\end{equation}
which means that one can proceed in the same way as in Sec.~\ref{SECfree},
through Eqs. (\ref{globconsn})-(\ref{f}).

The presence of interactions allows to give a more careful 
discussion of the local nature of Eq.~(\ref{param1}).
Naively, one might suspect that (\ref{param1}) may not be truly local 
because it involves the quantity $v_a^{\mu}(x_a)=j_a^{\mu}(x_a)/m_a$
obtained through a {\em nonlocal} integration in (\ref{eq6}).
To be more specific,
consider the case of $n=2$ entangled particles and assume that
$e_1=0$, so that $A^{\nu}(x)$ has a local influence
on the second particle $a=2$ only.
Assuming that $A^{\nu}(x)$ can be freely manipulated by an experimentalist,
can such a manipulation have a nonlocal influence on the first particle $a=1$? 
In our case, (\ref{eq6}) reduces to
\begin{equation}\label{eq6.1}
j_{1\mu}(x_1)=
\int_{\Sigma_2} dS^{\mu_2}_2  \, j_{\mu \mu_2}(x_1,x_2) .
\end{equation}
The quantity $j_{\mu \mu_2}(x_1,x_2)$ in (\ref{eq6.1}) depends 
on $A^{\nu}(x_2)$, so our question can be reduced to the following one: 
Does the left-hand side of (\ref{eq6.1}) depend on $A^{\nu}(x_2)$?
If it does, then the trajectory $X_1^{\mu}(s)$ determined by 
(\ref{param1}) depends on $A^{\nu}(x_2)$ as well, in which case
(\ref{param1}) is not truly local. So to show that (\ref{param1}) is truly local,
one needs to convince himself that the left-hand side of (\ref{eq6.1}) does not depend on 
$A^{\nu}(x_2)$. 

Fortunately, it is not difficult to understand why is that so. Indeed, this is very similar
to the well-known fact in nonrelativistic QM that local manipulations on the second
particle do not influence the marginal probability density
$\rho({\bf x}_1)=\int d^3x_2 \, \psi^*({\bf x}_1,{\bf x}_2) \psi({\bf x}_1,{\bf x}_2)$ 
of the first particle,
which is why EPR correlations cannot be used for superluminal signalling.
(In fact, the quantity (\ref{curn}) can even be thought of as a kind of relativistic generalization of
$\psi^*\psi$, in the sense that (\ref{curn}) reduces to $\psi^*\psi$ in the
nonrelativistic limit. More precisely, in this limit the largest component
of (\ref{curn}) is $j_{0\ldots 0} \simeq m_1\cdots m_n \, \psi^*\psi$,
while other components are negligible compared to this one.) 
In our relativistic case, the crucial observation is the fact that
the integral in (\ref{eq6.1}) does not depend on the choice of the integration 
hypersurface $\Sigma_2$, which is a consequence of (\ref{consnagain}). 
Consequently, if, for instance, the external field $A^{\nu}(x_2)$
is turned on at time $x^0_2=0$ before which 
$A^{\nu}(x_2)=0$, then one can choose $\Sigma_2$ to lie completely at
times before $x^0_2=0$, showing that the integral in
(\ref{eq6.1}) is the same as if the external field did not exist at all. 
Therefore, the left-hand side of (\ref{eq6.1}) does not depend on 
$A^{\nu}(x_2)$. 

Now consider the classical limit. This corresponds to a pure-phase wave function
\begin{equation}\label{wfclas}
\psi(x_1,\ldots,x_n) = {\rm const}\, e^{iS(x_1,\ldots,x_n)} . 
\end{equation}
Indeed, for such a wave function, the complex equation (\ref{KGA}) implies
a real equation
\begin{equation}\label{KGAclas}
-P_a^{\mu} P_{a\mu} +m_a^2=0 ,
\end{equation}
where
\begin{equation}
 P_{a\mu}\equiv \partial_{a\mu}S+e_aA_{\mu}(x_a) .
\end{equation}
The other independent real equation resulting from (\ref{KGA}) with 
(\ref{wfclas}) is $\partial_{a\mu}P_a^{\mu}=0$.
By dividing (\ref{KGAclas}) with $2m_a$ and summing over $a$, one gets 
the interacting generalization of (\ref{HJclasr}).
The solution $S(x_1,\ldots,x_n)$ of (\ref{KGAclas}) has a local form
\begin{equation}
 S(x_1,\ldots,x_n)=\sum_{a=1}^{n} S_a(x_a) ,
\end{equation}
so, with an appropriate choice of the normalization constant in (\ref{wfclas}), 
(\ref{param1}) reduces to
\begin{equation}\label{param1.cl}
 \frac{dX^{\mu}_a(s)}{ds} \stackrel{\rm traj}{=} -\frac{P_a^{\mu}(X_a(s))}{m_a} .
\end{equation}
From (\ref{param1.cl}) and (\ref{KGAclas}) one finds
\begin{equation}\label{proptime2}
 ds^2 \stackrel{\rm traj}{=} dX^{\mu}_a dX_{a\mu} ,
\end{equation}
which shows that $s$ is the proper time along the trajectories.
Eqs.~(\ref{KGAclas})-(\ref{proptime2}) are nothing but equations
of the classical relativistic Hamilton-Jacobi formalism for particles in a background
electromagnetic field $A^{\mu}(x)$. This shows that the theory has the correct 
classical limit.

The interaction with a background gravitational field can be introduced in a similar way.
Without going into details, we note that (\ref{consnagain}) generalizes to 
 \begin{equation}\label{consngrav}
 \nabla_{\mu_a}j^{\mu_1\ldots \mu_a \ldots \mu_n}=0 ,
\end{equation}
where $\nabla_{\mu_a}$ is the general-covariant derivative in curved spacetime
(see, e.g., \cite{weinberg}). Consequently, the integral (\ref{eq6}) in curved spacetime
does not depend on the choice of spacelike hypersurfaces, and one can proceed
in essentially the same way as without gravitation (except for the fact that
ordinary derivatives should be replaced by the general-covariant ones).

\section{Discussion}
\label{SECdisc}

In this paper, we have shown that nonlocality of QM  
can be made compatible with particle trajectories
$X^{\mu}_a(s)$ satisfying {\em local} and  {\em relativistic-covariant} laws,
where $s$ is equal (or locally related) to the proper time along the
trajectories. Instead of nonlocal laws for particle trajectories (as in Bohmian mechanics),
nonlocality is encoded in the hypothesis that the physical time is not $s$ but
a new parameter $s'$ (called modified proper time)
nonlocally related to $s$ via
\begin{equation}\label{essense}
 ds' \propto |\psi(x_1,\ldots,x_n)|^2 ds ,
\end{equation}
where $\psi(x_1,\ldots,x_n)$ is the standard relativistic many-time wave function.
The two proper times coincide in the classical limit (because in that limit
$|\psi|^2$ is a constant), which makes (\ref{essense}) consistent
with classical relativity.

We have explicitly discussed only the spin-0 case and we have not discussed
the effects of quantum field theory (QFT). Yet, our results can be
generalized to particles with spin and QFT in a straightforward manner,
by appropriate adaptation
of the formal developments presented in \cite{nikqft,nikSmatr}
in the context of Bohmian mechanics.
Let us briefly sketch how it works.
In the case of spin, the wave functions $\psi$ generalize to
wave functions $\psi_A$ with additional indices $A$, so expressions of the form
$\psi^*\cdots\psi$ (such as (\ref{curn}) or $|\psi|^2\equiv \psi^*\psi$) 
generalize to $\sum_A \psi^*_A\cdots\psi_A$.  
States in QFT are represented by wave functions which, in general, depend
on an infinite number of coordinates, so that even states which are not
eigenstates of the particle-number operator can be represented by wave functions.
Up to these generalizations, the rest of the theory is essentially the same
as that in Sec.~\ref{SECreal}.
Thus, our theory has a capacity to reproduce all predictions of
quantum theory.

We have not presented any independent argument for validity
of our main hypothesis (\ref{essense}), except for the fact that
this hypothesis reproduces the predictions of QM. 
Yet, the mere fact that such a simple hypothesis (combined with the appropriate local laws
for particle motions) is sufficient 
to reproduce the predictions of QM seems significant to us. 
Even if the theory described above does not describe the true reality
behind QM, at the very least it provides an example which explicitly demonstrates that 
quantum reality may have a different (and simpler) form of nonlocality
than suggested by the Bohmian interpretation. 
We believe that it significantly enriches the
general understanding of nonlocality and relativity in QM. 


\section*{Acknowledgements}

The author is grateful to T. Norsen for valuable discussions.
This work was supported by the Ministry of Science of the
Republic of Croatia under Contract No.~098-0982930-2864.

\appendix

\section{Conceptual issues: Internal time, external time, and nonlocal time}
\label{APP}

Even if the mathematics of this paper is straightforward,
a conceptually or philosophically inclined reader may have difficulties
to understand the novel concept of nonlocal time
on an intuitive level. The intention of this Appendix is to help the reader
to get a better conceptual understanding of it.
For that purpose, we find useful to first explain the concepts of internal and external time.

In classical mechanics, one usually thinks of time as a parameter ($t$ or $x^0$)
existing even without particle trajectories or any other mathematical curves
in space or spacetime. We refer to time existing
without any curve as {\em external time}. 
By contrast, the concept of proper
time $s$ is usually viewed as a parameter defined only on a 
curve in spacetime (usually a particle trajectory).
We refer to time defined only on a curve as {\em internal time}.

While proper time is typically viewed as an internal time and 
$t$ is typically viewed as an external time, in this section we want to explain
that these views can also be changed without contradicting any 
physical facts. There is a sense in which $t$ can be viewed as an
internal time, and there is also a sense in which 
proper time can be viewed as an external time.

The proper time $s$ along a curve in spacetime is given by
\begin{equation}\label{app1}
 ds^2 \stackrel{\rm curve}{=} dX^{\mu}dX_{\mu} .
\end{equation}
The right-hand side of (\ref{app1}) is defined only on a curve,
so $s$ in (\ref{app1}) is an internal time. But in that sense,
it is easy to see that the Newton time $t$ can also be viewed
as an internal time. For example, the trajectory $X^i(t)$
of a free nonrelativistic particle satisfies
$m \, dX^i(t)/dt \stackrel{\rm traj}{=} p^i$ with a constant momentum $p^i$. This implies
\begin{equation}\label{app2}
dt^2 \stackrel{\rm traj}{=} \frac{m}{2E} dX^i dX^i ,
\end{equation}
where the summation over the repeated index $i=1,2,3$ is understood and $E=p^i p^i/2m$.
Clearly, $t$ in (\ref{app2}) is an internal time in the same sense
in which $s$ in (\ref{app1}) is an internal time.

Does (\ref{app2}) imply that the Newton time $t$ is really an internal time?
One will say {\em no}, because (\ref{app2}) is {\em not the definition}
of $t$, but only corresponds to a special case (a free particle with a given momentum $p^i$).
On the other hand, (\ref{app1}) {\em is the definition} of $s$, so $s$ is really an internal time.
However, definitions are not facts given by nature. Instead,
definitions are conventions chosen by humans. Thus, if one {\em defines}
$t$ to be given by (\ref{app2}), then $t$ {\em is} the internal time.
Such an internal definition of $t$ is much more restrictive then the external one, but it may
be perfectly sensible if one wants to define time as the reading of a physical clock
(where the reading of the clock is identified with the position $X^i$ of its needle). 

The moral is that $t$ can be defined either as an internal time or an external time.
Both definitions are physical, but the external definition is more general. 
With the external definition of $t$, the space and time together can be viewed
as a 4-dimensional entity. (Such a 4-dimensional view makes sense even without relativity,
but relativity further reinforces the relevance of the 4-dimensional view.)
By contrast, with the internal definition of $t$, the world is better viewed as a 3-dimensional
entity. Yet, all these different views are nothing but different interpretations
of the same physical facts.

Just as $t$ can be interpreted as either internal or external time,
the same is valid for $s$. Even though it is common to view $s$ as an
internal time, it can be viewed as an external time as well.
The external view means that (\ref{app1}) is not the definition, but only a special
case of some more general theory
(such as \cite{niksuperlum} or the theory exposed in the present paper).
In the external view one deals with 5 independent 
parameters $s$, $x^0$, $x^1$, $x^2$, $x^3$, so one can think
of the world as a 5-dimensional entity. Such a view may seem
bizarre at first, but actually it is not much more bizarre than the
4-dimensional view of time and space in nonrelativistic physics.
Indeed, in the relativistic theory studied in \cite{niksuperlum},
the relativistic-scalar parameter $s$ is very much analogous to the nonrelativistic Newton time $t$.

The dimensionality of the ``world'' further increases when one considers
more than one particle. The configuration space for $n$ nonrelativistic
particles is a $3n$-dimensional space. The coordinates for this
space are $x^i_a$, for $a=1,\ldots,n$. This together with the external time
$t$ makes the total of $3n+1$ dimensions. Similarly,
a covariant formulation of the dynamics of $n$ relativistic particles requires
a $4n$-dimensional configuration space, with coordinates $x^{\mu}_a$.
This together with the external proper time $s$ makes the total of $4n+1$ dimensions.

With these multidimensional spaces at hand, we can study general coordinate transformations
on these spaces. On the $(3n+1)$-dimensional space, the general coordinate transformation
takes the form
\begin{eqnarray}\label{app3}
& x^i_a \rightarrow x'^i_a=f^i_a({\bf x}_1, \ldots , {\bf x}_n,t) , &
\nonumber \\
& t \rightarrow t'=f({\bf x}_1, \ldots , {\bf x}_n,t) . &
\end{eqnarray}
In particular, (\ref{e11}) is a coordinate transformation (\ref{app3}) with
\begin{eqnarray}\label{app4}
 & f^i_a({\bf x}_1, \ldots , {\bf x}_n,t) = x^i_a , &
\nonumber \\
& f({\bf x}_1, \ldots , {\bf x}_n,t) = \rho({\bf x}_1, \ldots , {\bf x}_n)t + {\rm const} . 
\end{eqnarray}
Similarly, on the $(4n+1)$-dimensional space, the general coordinate transformation
takes the form
\begin{eqnarray}\label{app5}
& x^{\mu}_a \rightarrow x'^{\mu}_a=f^{\mu}_a( x_1, \ldots , x_n,s) , &
\nonumber \\
& s \rightarrow s'=f(x_1, \ldots , x_n,s) . &
\end{eqnarray}
In particular, (\ref{e11r}) is a coordinate transformation (\ref{app5}) with
\begin{eqnarray}\label{app6}
 & f^{\mu}_a(x_1, \ldots , x_n,s) = x^{\mu}_a , &
\nonumber \\
& f(x_1, \ldots , x_n,s) = \rho(x_1, \ldots , x_n)s + {\rm const} . 
\end{eqnarray}

The physical meaning of general coordinate transformations is well understood
in the general theory of relativity, where the coordinate transformations
refer to the 4-dimensional spacetime. In particular, a coordinate transformation
of the form $t\rightarrow t'=f({\bf x},t)$ may correspond to a transformation
from a nonphysical coordinate time $t$ to a physical time $t'$. 
To some extent, the toy models and theories studied in the present
paper are similar
to the general theory of relativity, in the sense that our transformations
(\ref{app4}) and (\ref{app6}) are also coordinate transformations
to a physical time, but in more than 4 dimensions.
Since our multidimensional spaces involve an $n$-particle configuration space,
these transformations of the time coordinate are naturally viewed as
transformations nonlocal on the usual $(3+1)$-dimensional spacetime.
This is what we mean when we say that the physical time-parameters given by (\ref{app4}) and (\ref{app6}) are {\em nonlocal times}.

To conclude, time can be viewed either as an internal time or an external time,
depending on the context. The external view is more general and requires one dimension more
with respect to the internal view.
To understand the concept of nonlocal time, it is
useful to think of time as a time external to the configuration space.
The external view may be an unusual way of thinking when ``time'' refers
to the relativistic scalar time called ``proper time'', but is mathematically and physically
consistent.

\end{document}